# DISSIPATIVE QUADRATIC SOLITONS SUPPORTED BY LOCALIZED GAIN

Valery E. Lobanov[1], Olga V. Borovkova[1] and Boris A. Malomed[2]

[1]Russian Quantum Center, Skolkovo 143025, Russia

[2]Department of Physical Electronics, School of Electrical Engineering, Faculty of Engineering, Tel Aviv University, Tel Aviv 69978, Israel

## Abstract

We propose two models for the creation of stable dissipative solitons in optical media with the $\chi^{(2)}$ (quadratic) nonlinearity. To compensate spatially uniform loss in both the fundamental-frequency (FF) and second-harmonic (SH) components of the system, a strongly localized "hot spot", carrying the linear gain, is added, acting either on the FF component, or on the SH one. In both systems, we use numerical methods to find families of dissipative $\chi^{(2)}$ solitons pinned to the "hot spot". The shape of the existence and stability domains may be rather complex. An existence boundary for the solitons, which corresponds to the guided mode in the linearized version of the systems, is obtained in an analytical form. The solitons demonstrate noteworthy features, such as spontaneous symmetry breaking (of spatially symmetric solitons) and bistability.

## I. Introduction

Formation and stabilization of solitary waves is one of the central problems of nonlinear optics [1]. Self-trapped light waves confined in transverse directions or spatial solitons were found in a variety of physical settings. Solitary waves exist not only in conservative but also in dissipative materials where the self-localization is supported by the fundamental balance between the diffraction and self-focusing, as in lossless media, and between the loss and gain [2, 3]. Dissipative spatial solitons have been studied in different media, including lasers with saturable gain and absorption [2], systems where light evolution is governed by the cubic–quintic complex Ginzburg–Landau equation [3–9], semiconductor amplifiers [10].

A recently developed ramification of this topic is the study of trapped nonlinear modes in settings combining spatially localized gain ("hot spots") and uniform losses (see Ref. [11] for a brief review). Detailed theoretical analysis has been developed for diverse one- [12-26] and two- [27-32] dimensional realizations of models of this type, as well as for periodic distributions of the inhomogeneous gain and loss [33], and for both one- and two-dimensional discrete systems (lossy lattices) with the gain applied at a particular "hot site" of the lattice [34-36]. Related to this class of the models is the one with the spatially uniform linear gain and nonlinear loss whose strength grows from the center to periphery of the system at any rate faster than the distance from the center, which makes the dissipative solitons stable [37]. Also similar is the model in the form of the localized *PT dipole*, i.e., a merged [38] of separated [39] pair of mutually balanced point-like gain and loss elements embedded into the nonlinear medium, which supports a family of stable *PT*- (parity-time) symmetric solitons pinned to the pair.

The numerous works dealing with dissipative and *PT*-symmetric solitons supported by the localized gain ("hot spots") addressed media with the uniform [12-18, 20-22, 24-33, 37-39] or localized [19, 23, 34-36] Kerr nonlinearity. The objective of the present work is to extend the analysis of dissipative solitons supported by "hot spots" immersed into optical media featuring the quadratic, alias $\chi^{(2)}$, interactions, which is another generic type of the optical nonlinearity [40-43]. Previously, dissipative $\chi^{(2)}$ solitons were studied in models of spatially uniform optical cavities [44-46] and *PT*-symmetric systems [47, 48], but the option of using localized gain has not been explored yet.

An advantage offered by the $\chi^{(2)}$ media is based on the fact that they provide a very strong nonlinearity, when the mismatch between the fundamental-frequency (FF) and second-harmonic (SH) fields is small enough [40-43], hence the necessary propagation distance, i.e., the size of the experimental samples required for the observation of the solitons can be reduced to few centimeters. The strong nonlinearity also makes it possible to decrease the necessary power of the laser beams, which should be used to create the solitons – roughly speaking, from a multi-kW level to that of several Watt. However, a majority of perspective $\chi^{(2)}$-materials with high values of the quadratic nonlinearity (e.g., single-crystal organics and semiconductors, such as InAs, InSb, GaSb) feature excessive absorption at optical and near-infrared wavelengths. Therefore, to facilitate the soliton generation in such materials, one should compensate their intrinsic absorption, using, in particular, the concept of spatially localized gain. Besides providing the direct balance with losses, the localized gain may also serve as a means for steering nonlinear light beams (i.e., spatial optical solitons) [11].

We study the existence of the solitons in settings which combine the quadratic nonlinearity and localized gain acting on the single harmonic, either the SH or FF one, while the linear losses are present in both components. We demonstrate that such solitons may exist and be stable in a wide range of parameters. The structure of existence and stability domains may be rather complicated and complex effects, such as symmetry breaking and bistability, occur in them.

The rest of the paper is structured as follows. Two models, with the localized gain acting on either the FF or SH component, are formulated in Section II. Results of studies of dissipative solitons in the former system, both analytical and numerical, are reported in Section, which is followed by the presentation of results for the system with the gain built into the SH component in Section IV. The paper is concluded by Section V.

**II. The models**

First, we introduce the system with the "hot spot" acting on the FF. It is based on the system of coupled dimensionless equations for local amplitudes of the FF ($q_1$) and SH ($q_2$) waves in the spatial domain, under conditions for the type-I (degenerate) $\chi^{(2)}$ interaction [40-43] in the presence of the dissipation and localized gain:

$$\begin{cases} i\dfrac{\partial q_1}{\partial \xi} = -\dfrac{1}{2}\dfrac{\partial^2 q_1}{\partial \eta^2} - q_1^* q_2 \exp(-i\beta\xi) + i\gamma_1(\eta) q_1, \\ i\dfrac{\partial q_2}{\partial \xi} = -\dfrac{1}{4}\dfrac{\partial^2 q_2}{\partial \eta^2} - q_1^2 \exp(i\beta\xi) - i\gamma_2 q_2. \end{cases} \quad (1)$$

where $\eta$ and $\xi$ are transverse coordinate and propagation distance, respectively,

$$\gamma_1(\eta) = a\exp\left(-\eta^2/\eta_0^2\right) - \gamma_0, \quad (2)$$

with positive $\gamma_0$ and $a$, is the gain-loss profile at the FF with spatial width $\eta_0$, $\gamma_2 > 0$ represents the homogeneous loss at the SH, and $\beta$ stands for the wavenumber mismatch. Using the scaling invariance of the system, we fix $\eta_0 = 0.5$, and the generic results can be adequately presented, which is done below, for strength $\gamma_0 = 0.3$ of the background loss at the FF.

Note that, in the cascading limit [40-43], which, in the present case, corresponds to large values of $\beta$ and/or $\gamma_2$, the SH field can be eliminated,

$$q_2 \approx (\beta - i\gamma_2)^{-1} e^{i\beta\xi} q_1^2, \tag{3}$$

and the remaining equation for the FF,

$$i\frac{\partial q_1}{\partial \xi} = -\frac{1}{2}\frac{\partial^2 q_1}{\partial \eta^2} - \frac{\beta + i\gamma_2}{\beta^2 + \gamma_2^2}|q_1|^2 q_1 + i\gamma_1(\eta) q_1, \tag{4}$$

reduces to the hot-spot model with the cubic nonlinearity introduced in Ref. [13]. Here, our objective is to study the system in the properly $\chi^{(2)}$ regime.

The system with the localized gain acting at the SH is adopted in the following form, which can never be reduced to an effective counterpart with the cubic nonlinearity:

$$\begin{cases} i\dfrac{\partial q_1}{\partial \xi} = -\dfrac{1}{2}\dfrac{\partial^2 q_1}{\partial \eta^2} - q_1^* q_2 \exp(-i\beta\xi) - i\gamma_1 q_1 \\ i\dfrac{\partial q_2}{\partial \xi} = -\dfrac{1}{4}\dfrac{\partial^2 q_2}{\partial \eta^2} - q_1^2 \exp(i\beta\xi) + i\gamma_2(\eta) q_2 \end{cases}, \tag{5}$$

where $\gamma_1$ is the spatially uniform loss coefficient at the FF, while the spatial profile of the gain at the SH is taken as in Eq. (2), *viz.*,

$$\gamma_2(\eta) = a_2 \exp(-\eta^2/\eta_0^2) - \gamma_0. \tag{6}$$

By means of the rescaling we again set here $\eta_0 = 0.5$, while generic results are reported below for $\gamma_0 = 0.3$, $\gamma_1 = 0.5$.

### III. Solitons supported by the local gain applied at the fundamental-frequency harmonic

#### A. Analytical considerations

Stationary solutions for pinned states with real propagation constant $b$ are looked for as

$$q_1 = w_1(\eta)\exp(ib\xi), \quad q_2 = w_2(\eta)\exp(i(2b+\beta)\xi), \qquad (7)$$

where $w_{1,2}(\eta)$ are complex functions. The states are characterized by their FF and SH powers, $U_{1,2} = \int_{-\infty}^{+\infty} |q_{1,2}(\eta)|d\eta$, the total power (alias energy flow), $U = U_1 + U_2$, being the dynamical invariant of the system in the absence of loss and gain. In the presence of these terms, stationary solutions satisfy the power-balance condition,

$$\frac{dU}{d\xi} = 2\left[ a\int_{-\infty}^{+\infty} \exp\left(-\frac{\eta^2}{\eta_0^2}\right)|w_1(\eta)|^2 d\eta - \gamma_0 U_1 - \gamma_2 U_2 \right] = 0, \qquad (8a)$$

for the "hot spot" acting at the FF, and condition

$$\frac{dU}{d\xi} = 2\left[ a_2 \int_{-\infty}^{+\infty} \exp\left(-\frac{\eta^2}{\eta_0^2}\right)|w_2(\eta)|^2 d\eta - \gamma_1 U_1 - \gamma_0 U_2 \right] = 0, \qquad (8b)$$

for the "hot spot" acting at the SH.

Numerical solutions were constructed by means of the relaxation method.

In accordance with the general concept of linear gain-guided modes [49], the stationary pinned-mode solution to the linearized version of Eq. (1) exists at a single equilibrium value of the amplitude of the gain profile (2), $a = a_0$, which provides for the fulfillment of Eq. (8a), and at a single value of the propagation constant, $b_0$. In particular, an approximate solution to the linearized equations can be found for broad modes, with

$$\eta_0\sqrt{\gamma_0}, \; \eta_0\sqrt{-b_0} \ll 1. \qquad (9)$$

The *bulk part* of the respective solution, valid at $\eta^2 \gg \eta_0^2$, is

$$(q_1(\eta))|_{\text{bulk}} \approx Q_0 \exp\left(ib_0\xi - \sqrt{2(b_0 - i\gamma_0)}\,|\eta|\right), \quad q_2 = 0, \qquad (10)$$

with a negative propagation constant, $b_0$, and arbitrary amplitude $Q_0$, while the consideration of the solution in a vicinity of the "hot spot" amounts to relation

$$\Delta\left(\frac{\partial(q_1)_{\text{bulk}}}{\partial \eta}\right) \approx 2iQ_0 a \exp(ib_0\xi) \int_{-\infty}^{+\infty} \exp\left(-\frac{\eta^2}{\eta_0^2} - \sqrt{2(b_0 - i\gamma_0)}|\eta|\right) d\eta \qquad (11)$$

$$\approx 2iQ_0 a e^{ib_0\xi} \eta_0 \left[\sqrt{\pi} - \eta_0 \sqrt{2(b_0 - i\gamma_0)}\right],$$

where $\Delta(\partial q_1 / \partial \eta) = -2Q_0\sqrt{2(b - i\gamma_0)}$ stands for the jump of the $\eta$-derivative of the bulk solution (10) across the "hot spot". The substitution of this into Eq. (11) yields a complex algebraic equation:

$$a_0 \eta_0 \left(\sqrt{\frac{\pi}{2}} \frac{\sqrt{b_0 + i\gamma_0}}{\sqrt{b_0^2 + \gamma_0^2}} - \eta_0\right) = i, \qquad (12)$$

which can be reduced to the quartic equation for the propagation constant,

$$b_0 + \sqrt{b_0^2 + \gamma_0^2} = \frac{4}{\pi} \eta_0^2 \left(b_0^2 + \gamma_0^2\right), \qquad (13)$$

and an explicit expression for the gain strength:

$$a_0 = \frac{2}{\sqrt{\pi}\eta_0\gamma_0} \sqrt{\left(b_0^2 + \gamma_0^2\right)\left(b_0 + \sqrt{b_0^2 + \gamma_0^2}\right)}. \qquad (14)$$

In particular, for normalization $\eta_0 = 0.5$ adopted above, and typical value of the loss constant, $\gamma_0 = 0.3$, Eqs. (13) and (14) yield $a_0 \approx 1.222$, $b_0 \approx -0.445$, while the respective numerical findings are $a_0 = 1.323, b_0 = -0.389$. Finally, for very small values of $\gamma_0$, see Eqs. (5), Eqs. (13) and (14) can be simplified to explicit formulas:

$$a_0^3 = \gamma_0 / \left(\pi \eta_0^4\right), \quad b_0 \approx -(\pi/2)\eta_0^2 a_0^2. \qquad (15)$$

This result is compatible with the underlying condition (9) under the same condition which is adopted in the first inequality of (9), $\eta_0 \sqrt{\gamma_0} \ll 1$.

### B. Numerical results

We have found that numerically constructed dissipative $\chi^{(2)}$ solitons (see a typical example in Fig. 1) exist strictly at $a > a_0$ [here, $a_0$ is realized not necessarily as the approximate analytical result given by Eq. (12), but as the numerically found value admitting the existence of the linear gain-guided mode in the FF component]. This finding is quite natural, as the full model includes additional losses at the SH, the compensation of which requires an increase of the gain strength. Further, the existence and stability domains strongly depend on the sign of wavenumber mismatch $\beta$ (the stability was identified by means of systematic simulations of the perturbed evolution of the solitons, using the standard split-step fast-Fourier-transform algorithm). Figure 1 shows a typical soliton profile. Note, that stable soliton solutions exhibit a nontrivial phase distribution.

For $\beta > 0$, the range of the gain strength supporting stable solitons is significantly narrower than for $\beta < 0$. It is shown in Figs. 2 and 3 that, if the mismatch is positive there is a certain range of values of the gain coefficient where solitons do not exist even at $a > a_0$ [as it is shown at Fig. 2(*b*) at threshold points, the tangential line to the $U(a)$ curve becomes vertical, while $U$ remains finite]. For negative values of the mismatch, solitons exist at all values $a > a_0$ of the gain coefficient. Moreover, solitons do not exist for large values of the positive wavenumber mismatch [in Fig. 4 at the threshold value of $\beta$ the tangential line to the $U(\beta)$ curve becomes vertical, while $U$ remains finite], but they still exist for large absolute values of the negative mismatch, see Fig. 3 and Fig. 4. Note that this fact is opposite to the solitons' behavior as the function of wavenumber mismatch in conservative systems, where, while 1D solitons exist for any value of a positive mismatch, they are found only below a maximum absolute value of the negative mismatch. This finding can be explained in the above-mentioned cascading limit. Indeed, the positive mismatch in the so derived effective equation (4) corresponds to the self-focusing sign of the resultant cubic nonlinearity. However, it was shown earlier in Ref. [21] that in cubic medium stable gain-guided solitons exist at $a > a_0$ if the nonlinearity is defocusing. Note that the existence domain become wider with the growth of losses at the SH (see Fig.4).

The stability analysis has revealed that solitons are unstable when the gain strength exceeds a certain critical value, $a > a_{\mathrm{cr}}(\beta)$, which slightly depends on the mismatch, decreasing with its growth, see Fig. 3. The width of the stability domain increases with the growth of the SH loss rate, $\gamma_2$, all the solitons being unstable at $\gamma_2 = 0$.

**IV. Solitons supported by the local gain applied at the second harmonic**

Proceeding to the analysis of the model based on Eqs. (5) and (6), we first fix the mismatch $\beta$ and vary the gain strength $a_2$. It is thus found that stable solitons exists only at $\beta < 0$. Typical profiles of the solitons are shown in Fig. 5.

There are several branches of the dependence of the soliton's total power on the values $a_2$. The bottom branch (the red line marked "**b**" in Fig. 6) exists at $a_2 > a_{20} \approx 0.94528$, where $a_{20}$ corresponds to the gain-guided mode in the SH equation, with $U_1 \to 0$ but $U_2 \neq 0$, see Fig. 7. Naturally, $a_{20}$ does not depend on $\beta$, as, in terms of the linearized version of the SH equation in (5), $\beta$ reduces to an immaterial propagation-number shift. In the analytical form, $a_{20}$ can be predicted by Eqs. (13)-(15) with $\eta_0$ replaced by $\sqrt{2}\eta_0$, and $b_0$ replaced by the SG propagation constant. In particular, the approximation corresponding to Eq. (15) (very small $\gamma_0$) implies that $a_{20} = 2^{-2/3} a_0 \approx 0.63 a_0$, for the same values of $\gamma_0$ and $\eta_0$. As for the finite value of $U_2$ from which the branch originates, it is the one at which the parametric gain generated by the $\chi^{(2)}$ term in the first equation of system (5) provides for the compensation of the FF loss $\sim \gamma_1$.

Getting back to the numerical solutions of the full nonlinear system (5), we have found that they exist if the gain coefficient is smaller than a certain critical value (at threshold points the tangential line to the $U(a)$ curve becomes vertical, while $U$ remains finite). The existence domain shrinks when the mismatch decreases (compare different panels in Fig. 6) and disappears at small values of the mismatch (for the present parameters, it disappears, approximately, at $\beta < -4.35$, see Fig. 6(*c*) and Fig. 9). Solitons from this branch have a spatially symmetric shape. While $\beta$ increases, this branch become partially unstable, and, at values of $\beta$ large enough, stable solutions disappear. The intermediate branch (the green line in Fig. 6) is totally unstable. The upper branch (the black line marked "**u**" in Fig. 6) exists above a threshold values of the gain strength, which depends on $\beta$ (once again, at the threshold points the tangential line to the $U(a)$ curve becomes vertical, while $U$ remains finite). Solitons from this branch are spatially symmetric, actually staying stable below a point of the spontaneous symmetry breaking. At some value of the gain parameter, an asymmetric stable solution appears (the blue line marked "**as**" in Fig. 6), while the symmetric solutions become unstable. Asymmetric solutions are stable below some critical gain value (the blue circle in Fig. 6). At large values of $\beta$, stable solutions disappear. Interestingly, stable solitons from the upper and asymmetric branches may exist for small values of the gain strength, even at $a_2 < a_{20} \approx 0.94528$, i.e., below the above-mentioned value necessary for the existence of the gain-guided linear mode, see Eq. (13).

It is worth noting that there is a domain of bistability at Fig. 6(*a*), where two different symmetric solitons may exist for the same set of parameters. Profiles of such solitons are shown in top panels of Fig. 5. Also, for other values of $\beta$ a bistability domain for symmetric and asymmetric solutions may exist too, see Fig. 9.

Now, we fix the value of the gain parameter, $a_2$, and vary the mismatch, $\beta$, see Fig. 8. In this case, stable solutions from the bottom branch (the red line marked as "**b**", to the left from the red point) exist in a finite range of small negative values of the mismatch. Solutions from the upper branch (the black line marked as "**u**", on the left of the symmetry-breaking point) and asymmetric branches (the blue line marked as "**as**", to the right from the blue point) require larger values of the negative mismatch for their stability. As the gain decreases, the stability domain of the solitons from the upper and asymmetric branches shifts towards larger values of the negative mismatch, getting wider. For example, at $a_2 = 0.6$ the upper branch exists at $\beta < -12.9$, the symmetry breaking occurs at $\beta \approx -17.95$, and asymmetric solutions are stable at $\beta > -35.6$. However, such solitons imply high powers, which may be a problem in terms of their creation in the experiment.

Note that there is a domain of bistability at Fig. 8(*a*), where two different stable symmetric solitons may exist for the same set of parameters.

## V. Conclusions

We have introduced models which support stable dissipative solitons in media with the quadratic nonlinearity, into which a "hot spot" is embedded, with the linear localized gain acting either on the FF (fundamental-frequency) harmonic, or on the SH (second harmonic), with the spatially uniform linear losses present in both components. By means of numerical methods, we have found that $\chi^{(2)}$ solitons, pinned to the "hot spot", exist and are stable in wide ranges of parameters, in both systems. The existence boundary, which corresponds to the gain-guided modes in the linearized systems, was found in an approximate analytical form. The structure of the existence and stability domains may be rather complicated. Various phenomena, such as the spontaneous symmetry breaking of spatially symmetric solitons and bistability, have been revealed by the analysis.

It may be quite interesting to develop the analysis for 1D settings with pairs of the "hot spots", and for 2D settings too.

## VI. Acknowledgements

The work of Olga Borovkova was supported by an RQC Fellowship grant. B.A.M. appreciates hospitality of ICFO [Institut de Ciencies Fotoniques, Castelldefels (Barcelona), Spain], where a part of this work was done.

# Figure captions

**Fig. 1.** (color online) (*a*) Profiles of the absolute value of the FF (thick black line) and SH (thin red line) fields in the numerically found stable soliton supported by the local gain applied at the fundamental-frequency harmonic [model based on Eqs. (1) and (2)] at $\beta = 0$, $\eta_0 = 0.5$, $\gamma_0 = 0.3$, $\gamma_2 = 1.0$, $a = 2.0$. Profiles of the modulus (black dotted line), imaginary (thick blue line) and real parts (thin red line) of (*b*) FF and (*c*) SH fields. All quantities are plotted in dimensionless units.

**Fig. 2.** (Color online) (*a*) The soliton's energy flow (total power) vs. the gain coefficient at $\beta = -1.3$ (thick blue line), $\beta = 0$ (thin black line) and $\beta = 1.3$ (dashed line) for $\eta_0 = 0.5$, $\gamma_2 = 1.0$, $\gamma_0 = 0.3$. Colored dots indicate boundaries of the stability domains (solitons are stable to the left from this point). (*b*) Detailed dependence for the same parameters at $\beta = 1.3$. All quantities are plotted in dimensionless units.

**Fig. 3.** (Color online) Solitons' existence and stability domains at $\eta_0 = 0.5$, $\gamma_0 = 0.3$. Solitons exist to the right from the vertical red dotted line and below the solid lines. Stability domains are located to the left from the dashed lines of the same colour. All quantities are plotted in dimensionless units.

**Fig. 4.** (Color online) (*a*) The soliton total power (energy flow) and (*b*) propagation constant vs. wavenumber mismatch $\beta$ at $\eta_0 = 0.5$, $a = 2.0$, $\gamma_0 = 0.3$. The solitons are stable in the range of parameters displayed here. All quantities are plotted in dimensionless units.

**Fig. 5.** (color online) Profiles of the absolute value of the FF (thick black line) and SH (thin red line) fields of the solitons supported by the local gain applied at the second harmonic [model based on Eqs. (5) and (6)] at $\beta = -3$, $\eta_0 = 0.5$, $\gamma_0 = 0.3$, $\gamma_1 = 0.5$ for symmetric solitons at $a_2 = 0.98$ (*a*) from the bottom branch and (*b*) from the upper one, and (*c*) asymmetric soliton at $a_2 = 1.2$. Corresponding branches are shown at Fig. 6(*a*). All quantities are plotted in dimensionless units.

**Fig. 6.** (Color online) The soliton total power $U$ vs. gain coefficient $a_2$ for the solitons supported by the local gain applied at the second harmonic at $\gamma_0 = 0.3$, $\eta_0 = 0.5$, $\gamma_1 = 0.5$ for (*a*) $\beta = -3$, (*b*) $\beta = -1.75$ and (*c*) $\beta = -4.5$. The red line marked as "**b**" corresponds to the bottom branch, the black line marked as "**u**" – to the upper branch, the blue line marked as "**as**" – to the asymmetric branch, which emerges from the upper one through the spontaneous symmetry-breaking bifurcation, and the dashed green line – to the fully unstable branch. The blue dot indicates the boundary of the stability domain of the asymmetric branch (solitons are stable to the left from this point). All quantities are plotted in dimensionless units.

**Fig. 7.** (Color online) Solitons' powers $U_1$, $U_2$, $U$ vs. $a_2$ for the solitons from the bottom branch at $\gamma_0 = 0.3$, $\eta_0 = 0.5$, $\gamma_1 = 0.5$ for $\beta = -3$ (thick blue lines) and $\beta = -2$ (thin black lines). The solitons are stable in the range of parameters displayed here. All quantities are plotted in dimensionless units.

**Fig. 8.** (Color online) The solitons' total power $U$ vs. $\beta$ at $\gamma_0 = 0.3$, $\eta_0 = 0.5$, $\gamma_1 = 0.5$ for (*a*) $a_2 = 1.0$, (*b*) $a_2 = 1.3$ and (*c*) $a_2 = 0.85$. The red line marked as "**b**" corresponds to the bottom branch, the black line marked as "**u**" – to the upper branch, and the blue line marked as "**as**" – to the asymmetric branch. The blue dot indicates the boundary of the stability segment of the asymmetric branch (solitons are stable to the right from this point), and the red dot – the boundary of the stability segment of the bottom symmetric branch (solitons are stable to the left from this point). All quantities are plotted in dimensionless units.

**Fig. 9.** (Color online) Existence and stability domains at $\gamma_0 = 0.3$, $\eta_0 = 0.5$, $\gamma_1 = 0.5$. Red lines correspond to the bottom branch (branch "**b**" at Figs. 6 and 8). The existence domain is situated between the solid lines, these solitons being symmetric and stable to the left from the dashed line. Solitons from the upper branch (branch "**u**" at Figs. 6 and 8) exist above the solid black line, being symmetric and stable below the solid blue line, which indicates symmetry-breaking points. Asymmetric solitons (branch "**as**" at Figs. 6 and 8) exist above the solid blue line, and are stable below the dashed blue one. All quantities are plotted in dimensionless units.

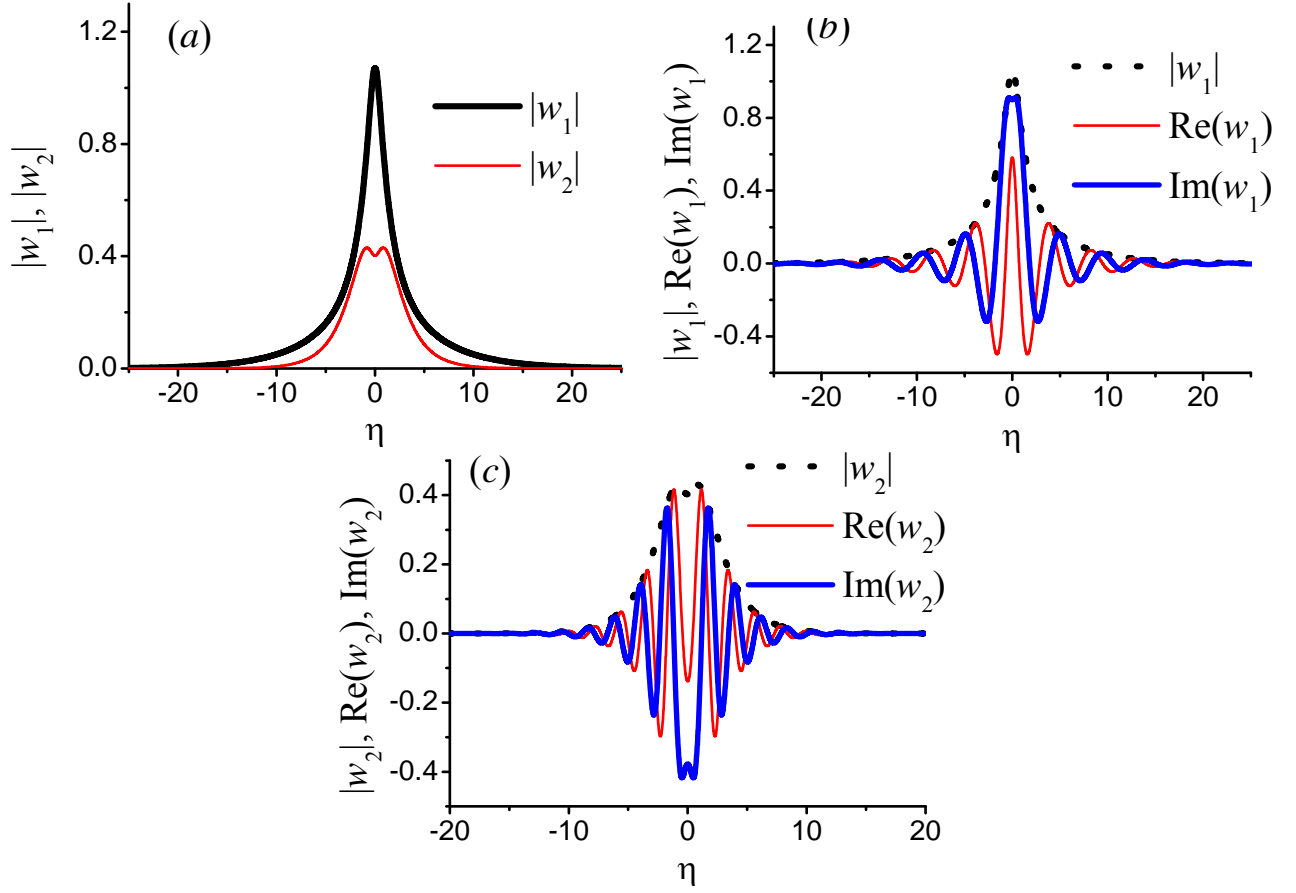

**Fig. 1.** (color online) (*a*) Profiles of the absolute value of the FF (thick black line) and SH (thin red line) fields in the numerically found stable soliton supported by the local gain applied at the fundamental-frequency harmonic [model based on Eqs. (1) and (2)] at $\beta = 0$, $\eta_0 = 0.5$, $\gamma_0 = 0.3$, $\gamma_2 = 1.0$, $a = 2.0$. Profiles of the modulus (black dotted line), imaginary (thick blue line) and real parts (thin red line) of (b) FF and (c) SH fields. All quantities are plotted in dimensionless units.

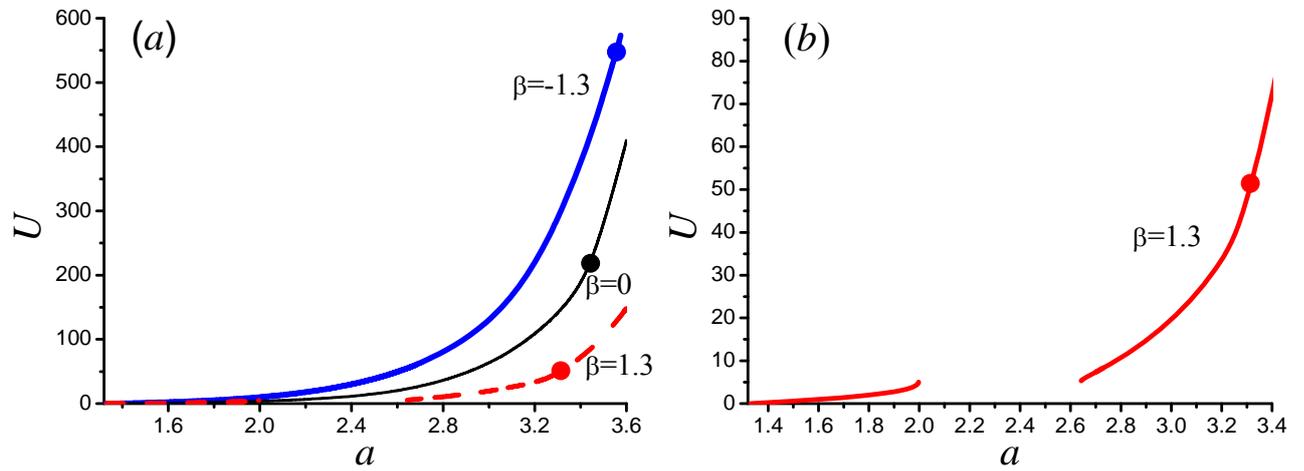

**Fig. 2.** (Color online) (*a*) The soliton's energy flow (total power) vs. the gain coefficient at $\beta = -1.3$ (thick blue line), $\beta = 0$ (thin black line) and $\beta = 1.3$ (dashed line) for $\eta_0 = 0.5$, $\gamma_2 = 1.0$, $\gamma_0 = 0.3$. Colored dots indicate boundaries of the stability domains (solitons are stable to the left from this point). (*b*) Detailed dependence for the same parameters at $\beta = 1.3$. All quantities are plotted in dimensionless units.

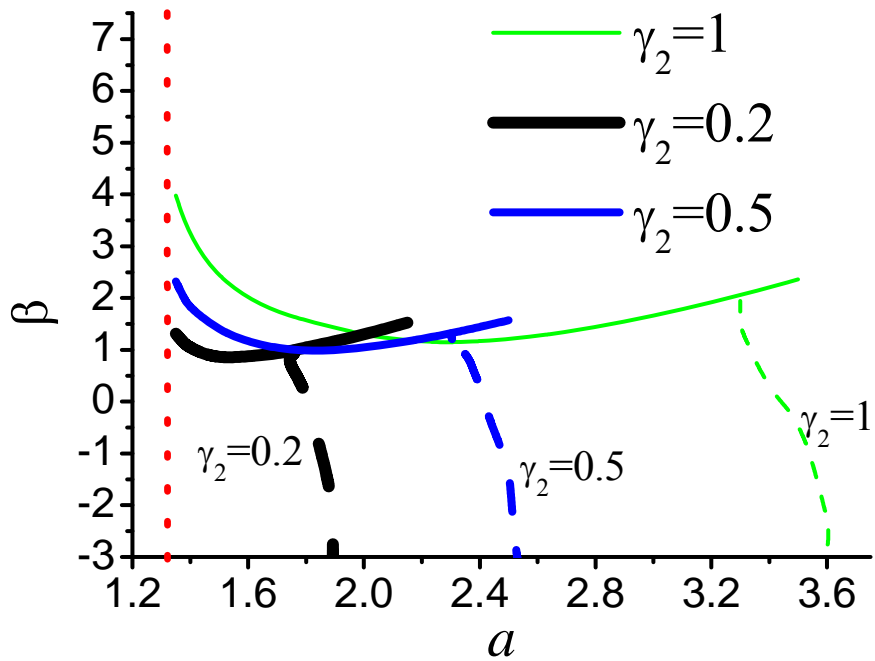

**Fig. 3.** (Color online) Solitons' existence and stability domains at $\eta_0 = 0.5$, $\gamma_0 = 0.3$. Solitons exist to the right from the vertical red dotted line and below the solid lines. Stability domains are located to the left from the dashed lines of the same color. All quantities are plotted in dimensionless units.

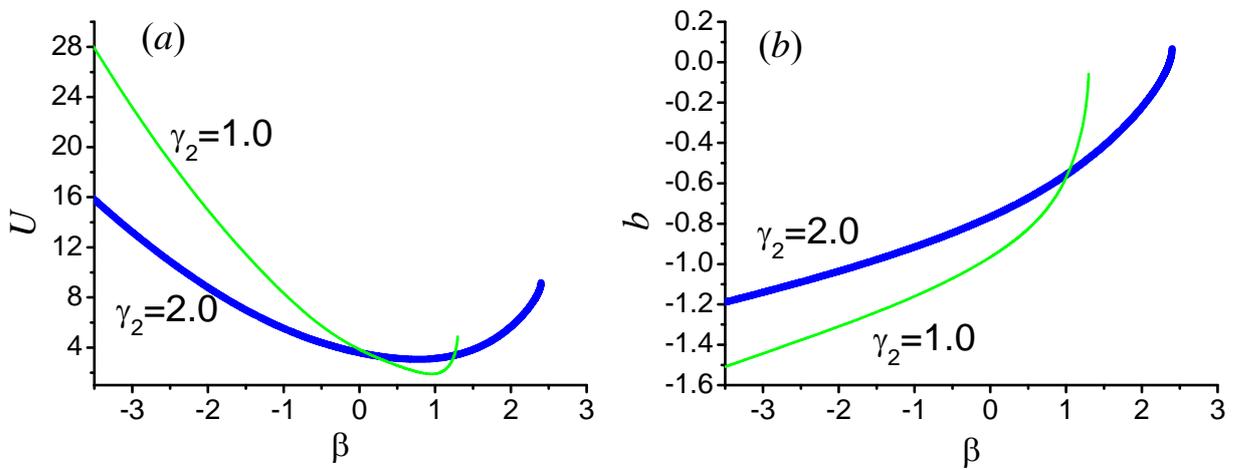

**Fig. 4.** (Color online) (*a*) The soliton total power (energy flow) and (*b*) propagation constant vs. wavenumber mismatch $\beta$ at $\eta_0 = 0.5$, $a = 2.0$, $\gamma_0 = 0.3$. The solitons are stable in the range of parameters displayed here. All quantities are plotted in dimensionless units.

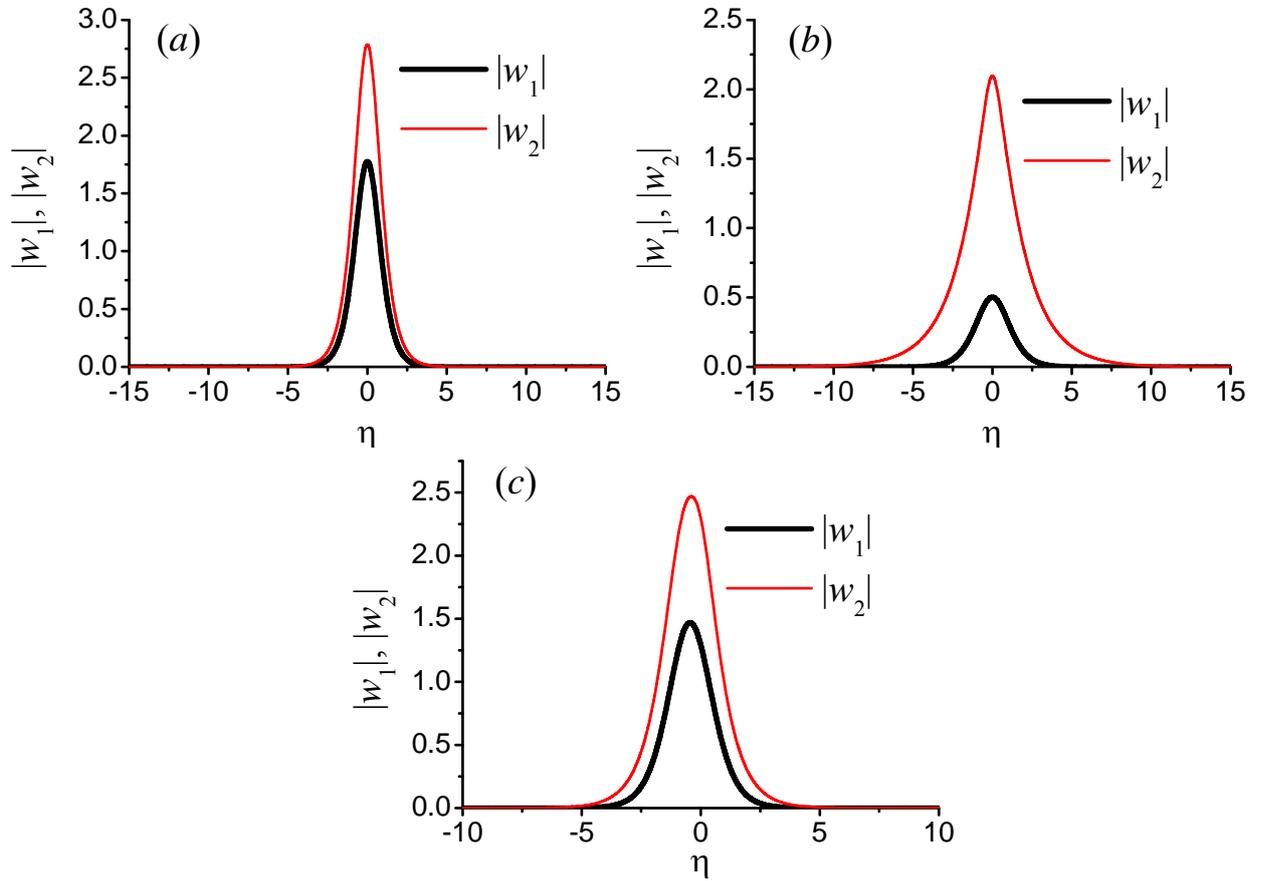

**Fig.5.** (color online) Profiles of the absolute value of the FF (thick black line) and SH (thin red line) fields of the solitons supported by the local gain applied at the second harmonic [model based on Eqs. (5) and (6)] at $\beta = -3$, $\eta_0 = 0.5$, $\gamma_0 = 0.3$, $\gamma_1 = 0.5$ for symmetric solitons at $a_2 = 0.98$ (*a*) from the bottom branch and (*b*) from the upper one, and (*c*) asymmetric soliton at $a_2 = 1.2$. Corresponding branches are shown at Fig. 6(*a*). All quantities are plotted in dimensionless units.

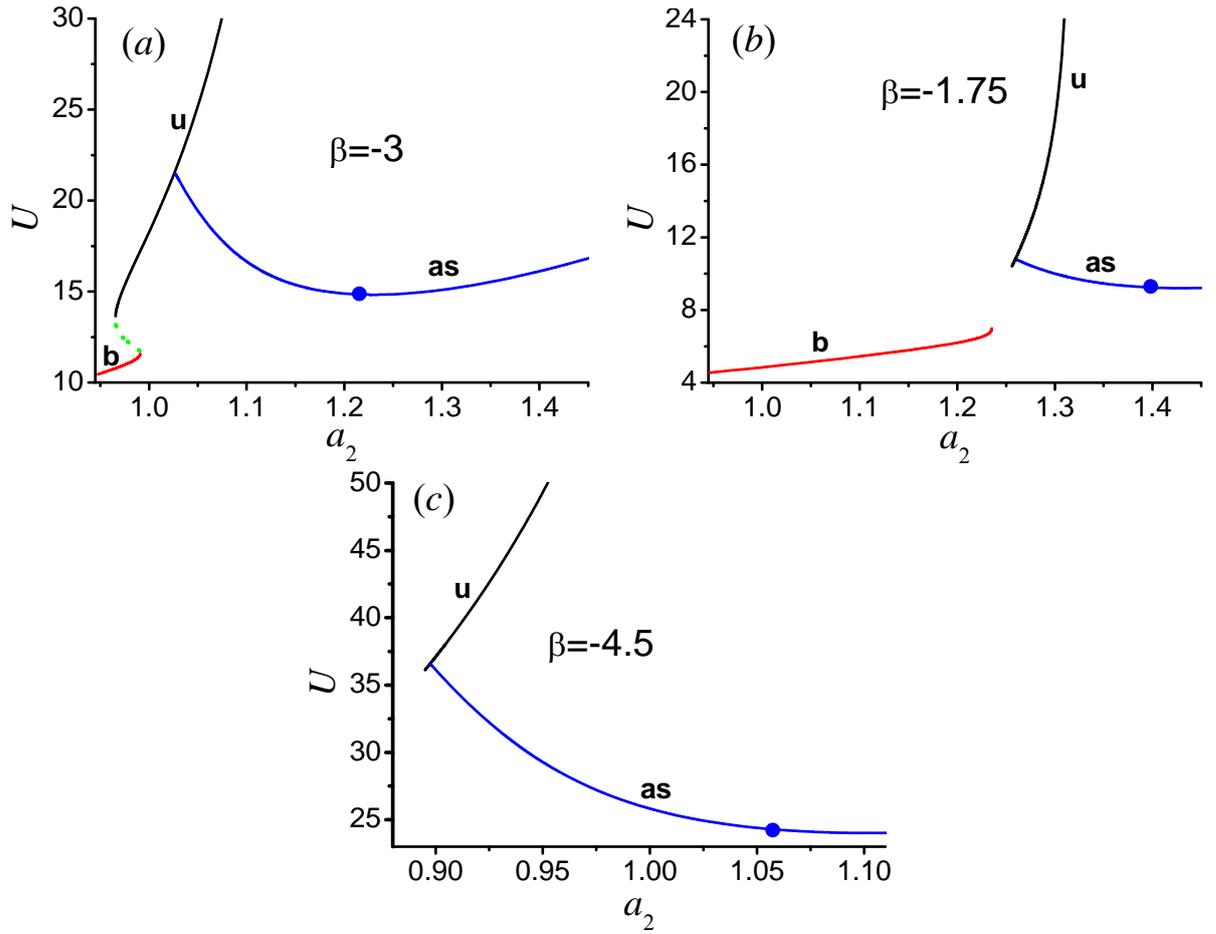

**Fig. 6.** (Color online) The soliton total power $U$ vs. gain coefficient $a_2$ for the solitons supported by the local gain applied at the second harmonic at $\gamma_0 = 0.3$, $\eta_0 = 0.5$, $\gamma_1 = 0.5$ for (*a*) $\beta = -3$, (*b*) $\beta = -1.75$ and (*c*) $\beta = -4.5$. The red line marked as "**b**" corresponds to the bottom branch, the black line marked as "**u**" – to the upper branch, the blue line marked as "**as**" – to the asymmetric branch, which emerges from the upper one through the spontaneous symmetry-breaking bifurcation, and the dashed green line – to the fully unstable branch. The blue dot indicates the boundary of the stability domain of the asymmetric branch (solitons are stable to the left from this point). All quantities are plotted in dimensionless units.

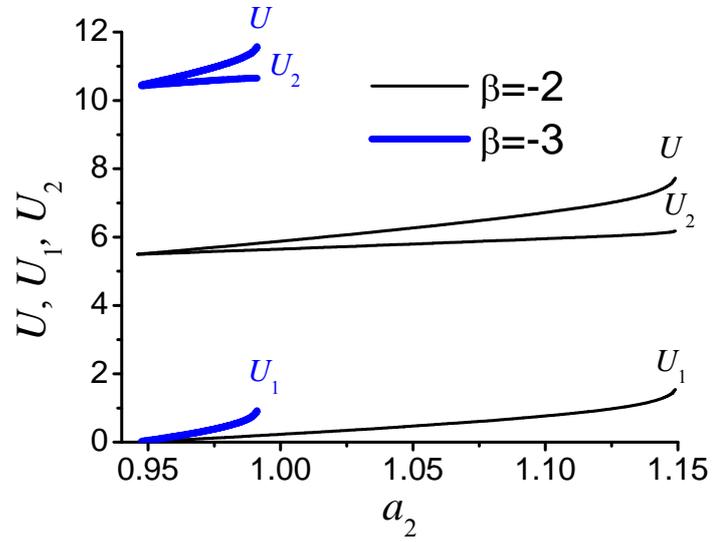

**Fig. 7.** (Color online) Solitons' powers $U_1$, $U_2$, $U$ vs. $a_2$ for the solitons from the bottom branch at $\gamma_0 = 0.3$, $\eta_0 = 0.5$, $\gamma_1 = 0.5$ for $\beta = -3$ (thick blue lines) and $\beta = -2$ (thin black lines). The solitons are stable in the range of parameters displayed here. All quantities are plotted in dimensionless units.

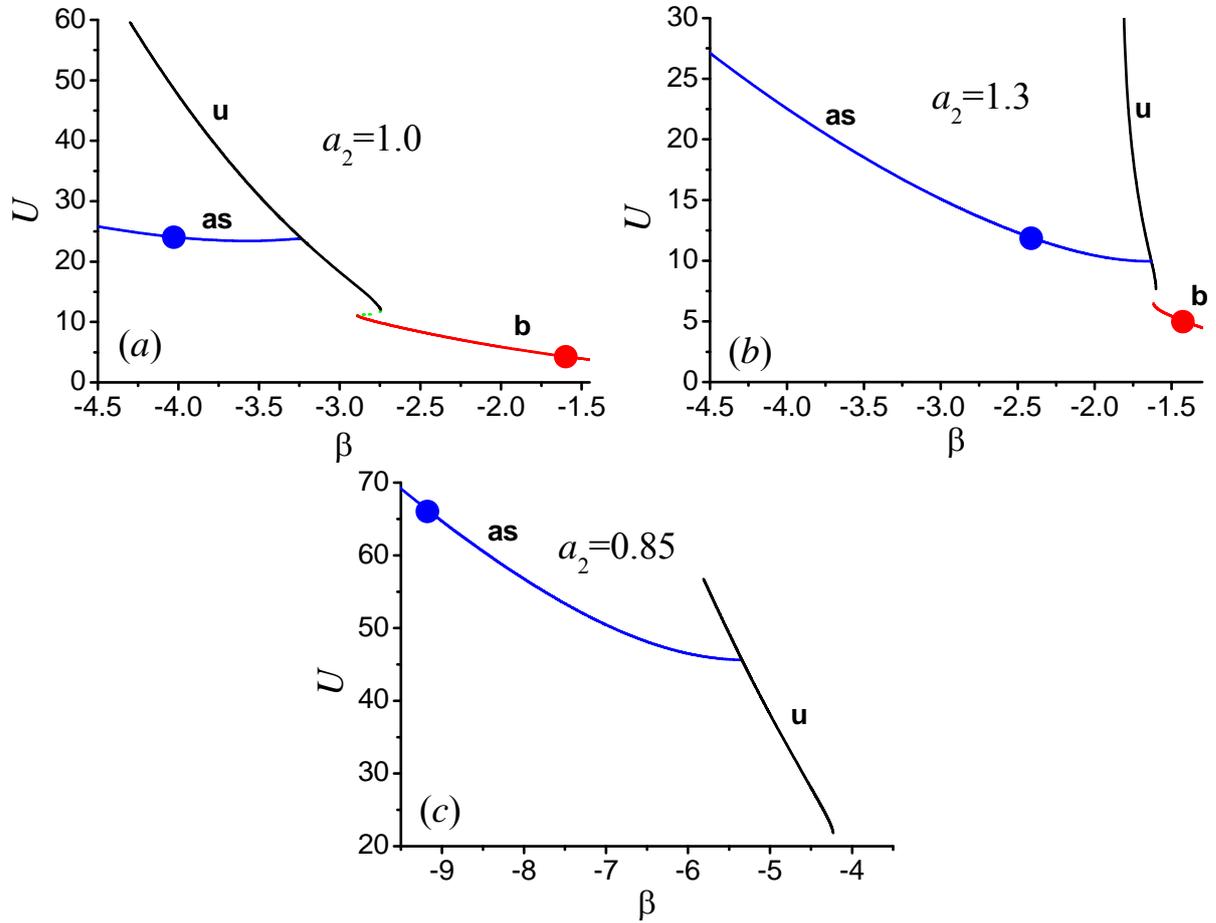

**Fig. 8.** (Color online) The solitons' total power $U$ vs. $\beta$ at $\gamma_0 = 0.3$, $\eta_0 = 0.5$, $\gamma_1 = 0.5$ for (*a*) $a_2 = 1.0$, (*b*) $a_2 = 1.3$ and (*c*) $a_2 = 0.85$. The red line marked as "**b**" corresponds to the bottom branch, the black line marked as "**u**" – to the upper branch, and the blue line marked as "**as**" – to the asymmetric branch. The blue dot indicates the boundary of the stability segment of the asymmetric branch (solitons are stable to the right from this point), and the red dot – the boundary of the stability segment of the bottom symmetric branch (solitons are stable to the left from this point). All quantities are plotted in dimensionless units.

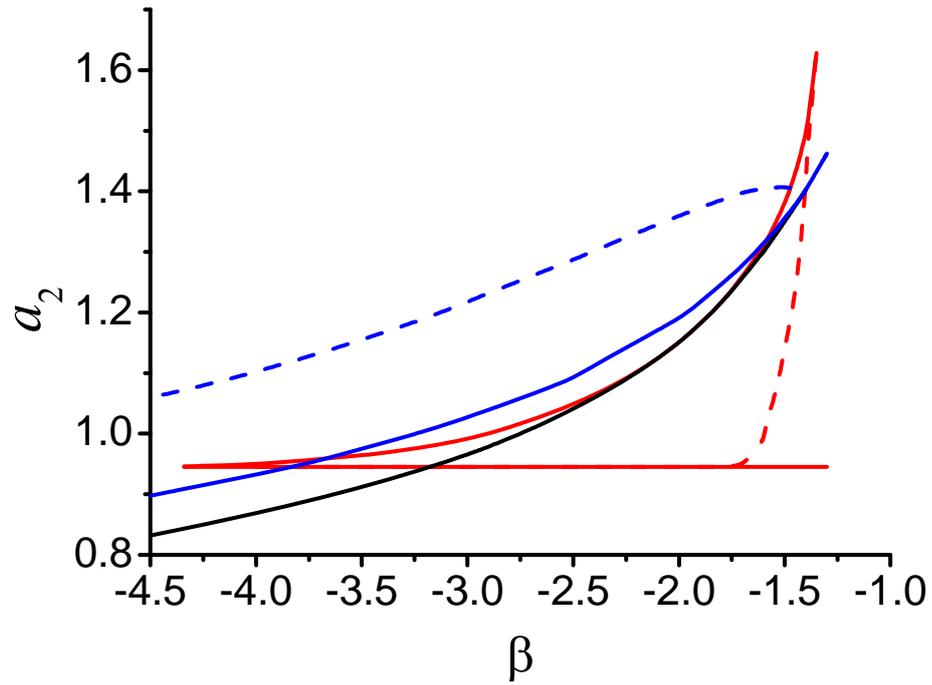

**Fig. 9.** (Color online) Existence and stability domains at $\gamma_0 = 0.3$, $\eta_0 = 0.5$, $\gamma_1 = 0.5$. Red lines correspond to the bottom branch (branch "**b**" at Figs. 6 and 8). The existence domain is situated between the solid lines, these solitons being symmetric and stable to the left from the dashed line. Solitons from the upper branch (branch "**u**" at Figs. 6 and 8) exist above the solid black line, being symmetric and stable below the solid blue line, which indicates symmetry-breaking points. Asymmetric solitons (branch "**as**" at Figs. 6 and 8) exist above the solid blue line, and are stable below the dashed blue one. All quantities are plotted in dimensionless units.